# Structural phase transitions and out-of-plane dust lattice instabilities in vertically confined plasma crystals

K. Qiao and T. W. Hyde

Center for Astrophysics, Space Physics and Engineering Research Baylor University, Waco, TX, 76798-7310, USA

ABSTRACT

The formation of plasma crystals confined in an external one-dimensional parabolic potential well is simulated for a normal experimental environment employing a computer code called BOX_TREE. Under appropriate conditions, crystals were found to form layered systems. The system's structural phase transitions, including transitions between crystals with differing numbers of layers and the same number of layers but different intralayer structures, were investigated and found to agree with previous theoretical and experimental research results. 1-2 layer transitions were examined in detail and shown to be caused by an out-of-plane lattice instability. Finally, growth rates for this out-of-plane lattice instability were obtained using the Box_Tree simulation with these results shown to agree with those obtained from analytical theory.

PACS number(s): 52.27.Lw, 52.35.Fp, 52.27.Gr, 61.50.Ah

1. INTRODUCTION

In a typical experiment on earth (under gravity), a plasma crystal is formed within a complex plasma when dust particles are levitated in the sheath region due to a balance between the gravitational and electrostatic force of a rf discharge [1-3]. The total external potential in this sheath region has been shown experimentally to approximate a parabolic potential well [4], thus it can be modeled as [5]

$$v(z) = \frac{\mu}{2}z^2. \qquad (1)$$

where $z$ is the particle height and $\mu$ is the parabolic coefficient.

In most cases, the particles will become negatively charged due to electron collection. However, this charge will be shielded by the ambient plasma thus the interaction between particles is best represented by a repulsive Yukawa potential defined by

$$v(r) = q\exp(-r/\lambda_D)/4\pi\varepsilon_0 r, \qquad (2)$$

where $q$ is the particle charge, $r$ is the distance between any two particles and $\lambda_D$ is the dust Debye length.

Structural phase transitions were first investigated by D. Dubin for a two dimensional One Component Plasma (OCP) system [6]. More recently, it has been proposed that an instability of the out-of-plane Dust Lattice Wave (DLW) [7-10] in a single-layer system may be the cause for the 1-2 layer transition. Such transitions have been observed in experiments on both OCP systems [11] and colloidal suspensions [12]. For plasma crystals, structural phase transitions have not yet been experimentally observed but have been investigated theoretically by H. Totsuji [5] employing a confined Yukawa system as a model. Totsuji established a phase diagram [5] for such a system characterized by two dimensionless parameters, $\kappa$ and $\eta$. The shielding parameter $\kappa$ is defined by

$$\kappa = \frac{a}{\lambda_D},  \qquad (3)$$

where $a$ is the mean interparticle distance as defined by $N_s = 1/\pi a^2$ with $N_s$ the surface number density in the $xy$ plane. $\eta$ is defined by

$$\eta = \frac{\mu}{4\pi q^2 N_s^{3/2}}. \qquad (4)$$

where q is the charge on the grain and µ is the parabolic coefficient as defined above.

In this research, the formation of a plasma crystal modeled by a vertically confined Yukawa system will be simulated using a numerical code called Box_Tree [9, 13-15]. The structural phase transitions will be investigated and compared with previous research results. The dispersion relation for the out-of-plane DLW will be obtained for a single-layer system, showing that the out-of-plane dust lattice instability appears at the point where the 1-2 layer transition starts [16]. This out-of-plane dust lattice instability will be examined with the instability growth rates calculated from Box_Tree simulations conducted for varying values of $\eta$. The dispersion relation for the out-of-plane DLW obtained using an analytical method [9] will also be used to calculate these growth rates and these analytical results will be compared to the simulation results.

## 2. STRUCTURAL PHASE TRANSITIONS

The dust particles in the plasma sheath are modeled as particles with a constant and equal charge $q = 3.84 \times 10^{-15} C$, equal mass $m_d = 1.74 \times 10^{-12} kg$ and radius $r_0 = 6.5 \mu m$. The interparticle interaction is assumed to be produced by a Yukawa potential with a Debye length $\lambda_D = 0.57 mm$ while the external potential is assumed to be parabolic in nature as shown by Eq. 1. The box size is set at $15 \times 15 \times 15 mm^3$, for a particle number of 600, thus the surface number density $N_s$ the mean distance $a$, and the shielding parameter $\kappa$ are equal to $2.67/mm^3$, $0.346 mm$ and $0.61$, respectively. Neutral gas drag is included with an Epstein drag coefficient [17] $\beta = 2.22 s^{-1}$. The boundary conditions in the XY directions are considered to be periodic since the box employed has a size much smaller than the size of the plasma crystal produced in a typical experimental environment. The boundary condition on the Z direction is assumed to be a closed boundary condition with particles hitting the top or bottom boundaries of the box reflected under an elastic collision.

Crystallization of the complex (dusty) plasma is simulated via the formation of an ordered crystal from an initially random distribution of particles placed in the box. Simulations were conducted for the range $0.48 \geq \eta \geq 0.0034$. As η decreases from 0.48 to 0.0034, the crystal transitions from a single-layer system to a two, three, four and five-layer system [16] with the system existing as a single layer system for $0.48 \geq \eta \geq 0.456$, a two-layer system for $0.336 \geq \eta \geq 0.072$, a three-layer system for $0.06 \geq \eta \geq 0.0216$, a four-layer system for $0.0204 \geq \eta \geq 0.0084$ and a five-layer system for $0.0072 \geq \eta \geq 0.0034$.

Between any two consecutive stages there is a structural phase transition. While within each stage, as η decreases the number of layers remains constant, the symmetry within each layer changes from a square to a hexagonal (triangular) lattice. Between these symmetries, there is a transition stage where the lattice exhibits a complex structure showing a mixture of both square and hexagonal symmetry [16]. All of the above agree with previous theoretical [5, 6] and experimental [11, 12] results.

As can be seen in Fig. 1, for both square and hexagonal symmetry the vertical projection of a particle within one layer always falls onto the center of the lattice cell in the adjacent layer. As can also be seen in Fig. 1(b) and (d) for hexagonal lattices, particles within the third layer are primarily aligned vertically with particles in the first layer. This is characteristic of an hcp lattice, in which particle positions within a hexagonal lattice plane repeat themselves every other plane; thus the planes are ordered as ABABAB ….

In fact, the fcc phase, for which the lattice planes are ordered as ABCABC …, is the thermodynamically preferred state for a layered hexagonal structure since its Helmholtz free energy is smaller. However, for a Yukawa interaction with short range, the difference between the free energies of the fcc and hcp phases is very small, and local fluctuations of the particle density might result in a transition from the fcc to the hcp phase [18]. This is presumably why the hcp rather than the fcc structure was observed.

No vertical alignment was seen in any of the simulations. This is to be expected since the vertical alignment observed in plasma crystal experiments on earth is created by the ion flow wake effect while the system considered by this research is a pure Yukawa system [16].

The relative thickness of the system was also investigated as a function of $\eta$ [16]. A general increase in thickness as $\eta$ decreases and discontinuities in the d-$\eta$ function corresponding to the stepwise transitions in the number of layers N can be clearly seen in Fig. 2. Also shown is the dependence of the intralayer structures on $\eta$. It can be seen that the structural phase transitions and the d-η function are in agreement quantitatively with Totsuji's predictions [5].

## 3. 1-2 LAYER TRANSITION

As can be seen from the d-η function (Fig. 2) the 1-2 layer transition is characterized by a dramatic increase in overall system thickness. In Ref. [6], it was shown that for an OCP, a single layer system would first go through a 1-3 layer transition caused by the out-of-plane lattice instability. As η decreases, the 3-layer system will then change to a 2-layer system via a first order phase transition. To examine the 1-2 layer transition (which begins when $0.432 \geq \eta \geq 0.408$) in greater detail, Box_Tree simulations were conducted with a fine adjustment of η values around the transition point, for the range $0.444 \geq \eta \geq 0.418$. The corresponding d-η function is shown in Fig. 3(a). As can be seen, the critical value of η signifying the start of the 1-2 layer transition falls between $0.430 \geq \eta_{critical} \geq 0.427$.

The dispersion relation for the out-of-plane lattice wave has recently been derived using an analytical method [9] as

$$\omega^2 + i\beta\omega = \frac{\mu q}{m_d} - 2 \sum_{m,n \neq j,l} \frac{g_{mn}^{00}}{m_d} \sin^2\left(\frac{kx_0^{mn}}{2}\right). \qquad (5)$$

The condition, which must be met for instability in the out-of-plane lattice wave to occur, is for Eq. 5 to have a complex solution for ω. When this happens, the dispersion relation of the wave propagating parallel to the prime translation vector intersects the $\omega = 0$ axis [6].

Using Eq. 5, the dispersion relation for the out-of-plane lattice wave propagating parallel to the prime translation vector was found [16] to intersect the $\omega = 0$ axis when

$\mu \leq 3.09 \times 10^{-9} \, kg/s^2$ (which corresponds to $\eta \leq 0.4274$). The solid line in Fig. 3(b) shows this theoretical dispersion relation when η is equal to the threshold value of 0.4274.

A Box_Tree simulation was also run for this value of η and then employing the method in [9], the dispersion relation of the out-of-plane lattice wave was represented by an intensity graph for the particle velocities in k-ω space (Fig. 3(b)). As can be seen, the two methods agree well (Fig. 3(b)) with both showing the threshold value of η for the out-of-plane lattice instability as 0.4274. Comparing this value of η with the critical value of η signifying the start of the 1-2 layer transition, $0.430 \geq \eta_{critical} \geq 0.427$, it is verified that for a vertically confined Yukawa system the 1-2 layer transition starts at the point where an instability of the out-of-plane lattice wave (or out-of-plane lattice instability) appears [16].

There is some evidence that the 3-layer structure caused by the out-of-plane lattice instability as predicted in Ref [6] might be visible at the center of the layered system for $\eta = 0.4056$ (Fig. 4). The fact that it only appears at the center of the layered system is presumably due to the artificial periodic boundary conditions employed in the simulation.

## 4. THE OUT-OF-PLANE LATTICE INSTABILITY

The growth rate of the out-of-plane lattice instability is described by the imaginary part $\omega_i$ of the complex frequency ω. Eq. 5 was solved for $\eta \leq 0.4274$ and the complex solution for ω was obtained. The real ($\omega_r$) and imaginary ($\omega_i$) parts of this solution as a function of the wave number k for $\eta = 0.420$, 0.408, 0.396 and 0.384 are shown in Fig. 5.

From Fig. 5(b), it can be seen that the maximum growth rates occur for waves with wave number $k = 6.4 mm^{-1}$ and $k = 12.8 mm^{-1}$ and that only waves whose wave numbers fall within a specific range around these two values will have a positive growth rate. All other waves exhibit a small constant negative growth rate, which is caused by damping due to neutral gas drag. Both the growth rates and the range of allowable wave numbers for growing waves increase as η decreases.

To verify these analytical predictions, Box_Tree simulations for time-dependent η values were conducted. The simulations were started with a η value of 0.432, which is greater than the threshold value of 0.4274, thus insuring there would be no out-of-plane lattice instability or 1-2 layer transition. After 65 seconds the system stabilized as a single layer crystal. At this time η was changed to a value below the threshold value of 0.4274, at which point out-of-plane lattice instabilities appeared.

These instabilities, or growing waves were then detected and investigated by analyzing the output data from Box_Tree. Particle motion was tracked for approximately 1 second after the change of η with output data files created every 0.001-second with a total of 1000 data files obtained. Once this data was collected, the x-y plane of the box for each set of particles was divided into bins. For each data file the average vertical displacement from equilibrium for the particles within each bin was determined, yielding a matrix of positions with column number equal to the number of bins and line number equal to the number of data files. Since each file was collected at a specific time, and particles within each bin had a specific x coordinate, this position matrix yields particle displacements, which are both time (t) and position (x) dependent. This is the same method employed in [9] except that vertical displacements instead of velocities are considered. Since the simulation results were to be compared with growth rates for waves propagating parallel to the prime translation vector, the bins were chosen perpendicular to it [9].

A Fourier transformation of this matrix about x (Eq. 6) yields a new matrix representing particle displacement in k-t space.

$$Z_{k,t} = 2/TL \int_0^L z(x,t) \exp[-ikx] dx \qquad (6)$$

Since this matrix was obtained from the random particle motion on the vertical direction, it represents the time dependence of the magnitude of the thermally excited out-of-plane lattice waves for various wave numbers. As shown in Fig. 6, this data can be represented by an intensity graph in k-t space, where the magnitude of the waves is given by pixel brightness.

Fig. 6 shows an intensity graph obtained from a simulation for $\eta = 0.408$ after the system stabilized as a single layer crystal. It can be seen that for waves with k values of approximately 6 and 13, the magnitude increases with time, while waves with other k values do not show any apparent growth. To determine whether the range of k values for growing waves agrees with that found analytically, the theoretical growth rate curve obtained for $\eta = 0.408$ is superimposed on the right side of the graph. As can be seen, the range of wave numbers for growing waves is in good agreement with the analytical results.

The behavior of the waves can be seen more clearly by plotting the magnitude of a specific wave with a particular wave number as a function of time. This is shown for $k = 2.36$ and $k = 7.07$ in Fig. 7 (a) and (b) respectively. As can be clearly seen, for $k = 2.36$, the magnitude of the wave is small (below $1.2 \times 10^{-4} m$) and random, while for $k = 7.07$, the magnitude exhibits a smooth increase for the first $0.5s$ and then stabilizes with a value of approximately $5.0 \times 10^{-4} m$ for the second $0.5s$. It can also be seen that the increase of magnitude is almost exponential, as would be expected from instability theory. Thus, the growth rate can be determined by employing an exponential fit to the increasing portion of the curve. Doing so shows the growth rate to be $\omega_i = 6.30 s^{-1}$ as compared to that found from analytical theory, where it was $6.86 s^{-1}$.

Growth rates for out-of-plane lattice instabilities (growing waves) for $\eta = 0.420$, $0.396$ and $0.384$ have also been determined using this method and are shown in Fig. 8 along with those for $\eta = 0.408$. Only growth-rates for waves with wave numbers of approximately $k = 6.4$ are calculated since the magnitude of waves with $k = 12.8$ are too small to show any reasonable increase. Also, only data showing a smooth increase in wave magnitude within the first half second are used. The analytical results for these η values are also shown in Fig. 8. As can be seen, the analytical prediction that the growth rate increases with decreasing η is verified by the simulation data. Although, the average percentage difference between the simulation and the analytical results is approximately 20%, calculations show that this difference most likely arises from round-up error of the simulations. Another possible error comes from the fact that for small η ($\eta < 0.400$), the system stabilizes after approximately 0.5 second, exhibiting a rough two-layer shape. Since the analytical prediction is made based on the assumption that the system is a single-layer system, a deviation from the analytical prediction is to be expected.

## 5. CONCLUSIONS

The structural phase transitions within a plasma crystal modeled as a vertically confined Yukawa system, including both transitions between different numbers of layers and intralayer structures were simulated using the Box_Tree code. The generated results agree with previous theoretical and experimental results.

Box_Tree was also employed to investigate both the 1-2 layer transition and the out-of-plane instability for a vertically confined Yukawa system. The critical value of $\eta$ for the onset of the 1-2 transition and the $\eta$ value where the out-of-plane lattice instability begins to appear were both determined by an analysis of the out-of-plane lattice wave using both analytical and simulation methods. The values of $\eta$ obtained were shown to agree with one another thus showing that for a Yukawa system, the transition starts at the point where the out-of-plane lattice instability appears. The resulting 3-layer system caused by this instability was observed at the center of the system. The out-of-plane lattice instability was itself examined and growth rates obtained for systems with differing $\eta$ values again using both analytical and simulation methods. The growth rate values obtained were found to agree and both methods showed that for all $\eta$ values investigated, the highest growth rates occurred for waves with wave number $k = 6.4 mm^{-1}$ and $k = 12.8 mm^{-1}$. Only waves whose wave numbers fall within a $\eta$ dependent range around these two values can become instabilities, or growing waves.

FIG. 1. Top view of a vertically confined Yukawa system when (a) $\eta = 0.0204$, (b) $\eta = 0.0084$ (four-layer system), (c) $\eta = 0.0066$ and (d) $\eta = 0.0034$ (five-layer system). The asterisks, circles and triangles represent particles in the first, second and third layers, respectively.

FIG. 2. The system's relative thickness $d/a$ as a function of the characteristic parameter η. The intralayer structure with square, triangular and complex symmetries are represented by squares, triangles and circles respectively.

FIG. 3 (color online). (a) The d-η function for $0.444 \geq \eta \geq 0.418$ obtained from Box_Tree simulations and (b) the dispersion relation of the out-of-plane lattice wave propagating parallel to the prime translation vector obtained using both the analytical method and a Box_Tree simulation for the vertically confined Yukawa system ($\eta = 0.4274$).

FIG. 4. Side view of the vertically confined Yukawa system when $\eta = 0.4056$.

FIG. 5. $\omega_r$ and $\omega_i$ as functions of wave number k for $\eta = 0.420$, 0.408, 0.396 and 0.384.

FIG. 6 (color online). An intensity graph showing the time dependence of the magnitude of thermally excited out-of-plane lattice waves for various wave numbers for $\eta = 0.408$.

FIG. 7. Magnitude of the out-of-plane lattice wave as a function of time for $k = 2.36 mm^{-1}$ and $k = 7.07 mm^{-1}$ ($\eta = 0.408$).

FIG. 8. Growth rates of the out-of-plane lattice instabilities from simulation (shown by the symbols) and analytical theory (shown by the lines).

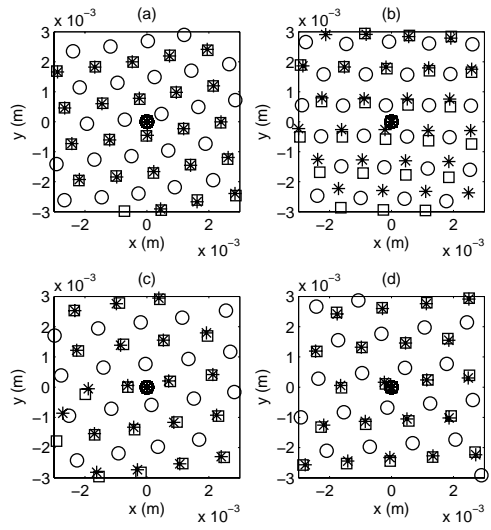

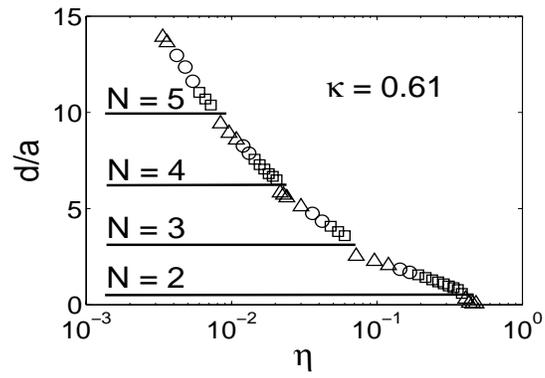

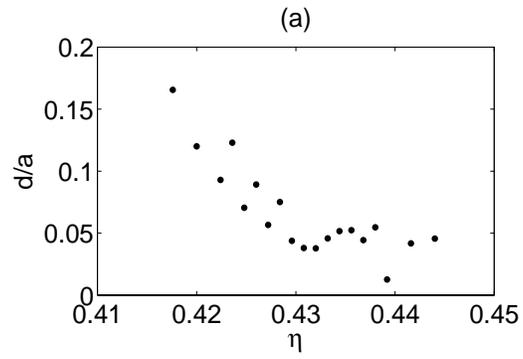

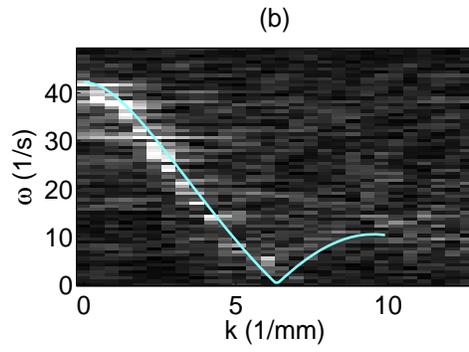

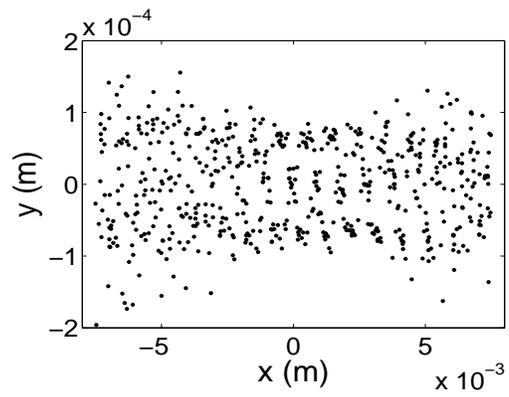

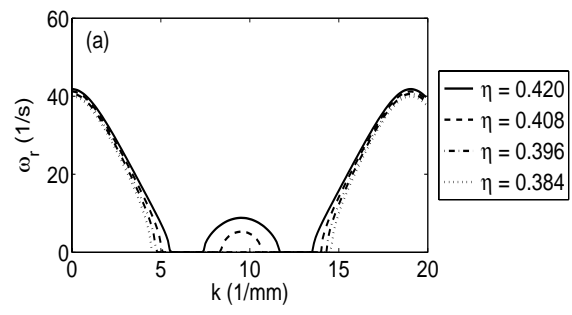

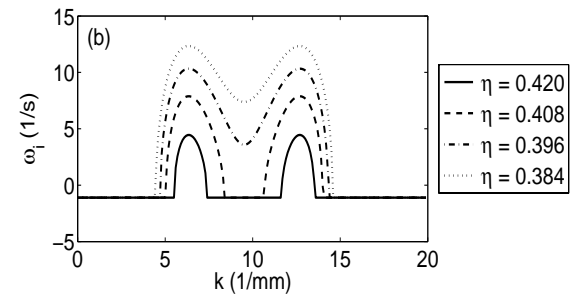

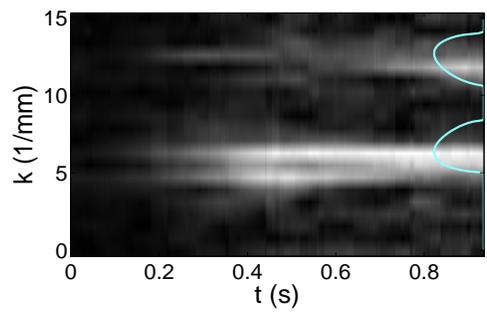

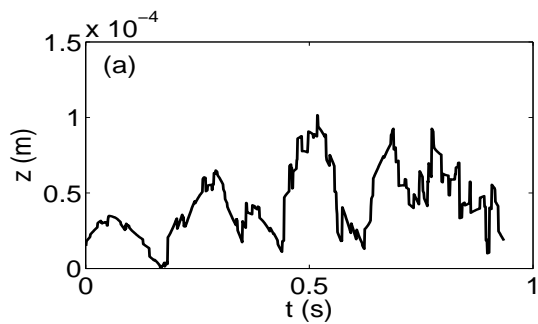

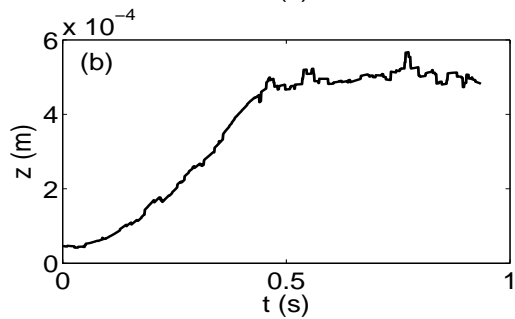

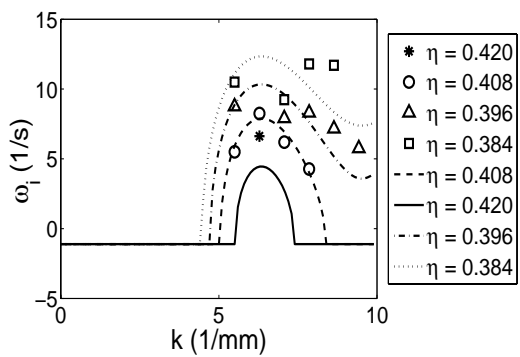